# Electronic and Level Statistics Properties of Si/SiO$_2$ Quantum Dots

I. Filikhin, S.G. Matinyan, B.K. Schmid and B. Vlahovic

Department of Physics, North Carolina Central University,

1801 Fayetteville Street, Durham, NC 27707, USA,

e-mail: ifilikhin@nccu.edu; fax: 919 530 7472

## ABSTRACT

Spherical shaped Si quantum dots (QDs) embedded into the SiO$_2$ substrate are considered in the single sub-band effective mass approach. Electron and heavy hole sub-bands are taken into account. Nonparabolicity of the Si conduction band is described by the energy dependence of electron effective mass. Calculations of low-lying single electron and hole energy levels are performed. For small sizes QD (diameter $D \leq 6\,\mathrm{nm}$) there is a strong confinement regime when the number of energy levels is restricted to several levels. The first order of the perturbation theory is used to calculate neutral exciton recombination energy taking into account the Coulomb force between electron and heavy hole. The PL exciton data are reproduced well by our model calculations. We also compare the results with those obtained within model [2]. For weak confinement regime (size $D \geq 10$ nm), when the number of confinement levels is limited by several hundred, we considered the statistical properties of the electron confinement. Distribution function for the electron energy levels is calculated and results are discussed.

## 1. Introduction

The Si/SiO$_2$ heterostructure has wide perspectives in the applications to the various optoelectronic nanodevices [1]. Theoretical study of the energy spectra properties of carriers confined in the Si/SiO$_2$ QDs is important for understanding the related electronic processes. A number of works devoted to the evaluation of the electron/hole energy levels [2] have been performed under the effective mass approximation using the envelope functions. In the present work we also used this method. However our approach is supplemented by energy dependence of electron effective mass [3] which is important for small size nanocrystals (diameter $D \leq 6$ nm). The non-parabolic effect taken into account under this approximation leads to the shift of the energy levels and changes of the electron/hole effective mass in the Si nanocrystals relative to the parabolic approach. In this work we evaluate values of these changes and compare our results for PL spectrum with previous results [2] and the experimental data. In small size QDs the several confinement levels exist (strong confinement regime). In this case the Coulomb force between electron and heavy hole can be taken into account by using the first order perturbation theory. The approximation is used for calculation of the neutral exciton recombination energy.

For weak confinement regime (QD size $D \geq 10$ nm), when the number of confinement levels is limited by several hundred we consider the statistical properties of the electron neighboring levels in QD. Evidence for the level repulsion of the close energy electron spectrum is demonstrated. As the reason for the onset of such behavior we consider two factors. The first is geometry of QD shape and the second one is effective mass anisotropy ($m_{0,1\parallel}/m_{0,1\perp} = 4.8$ for Si). In the present work we explore which factor is important for such behavior.

## 2. Model

A Si/SiO$_2$ heterostructure is modeled using a kp-perturbation single sub-band approach with energy dependent quasi-particle effective mass [3]. The energies and wave functions of a single carrier in a semiconductor structure are solutions of nonlinear Schrödinger equation:

$$\left( -(\nabla, \frac{\hbar^2}{2m^*(r,E)} \nabla) + V(r) - E \right) \psi = 0$$
, (1)

where $V(r)$ is the band gap potential, proportional to the energy misalignment of the conduction (valence) band edges of the Si QD (index 1) and the SiO$_2$ substrate (index 2). $V(r) = V_c$ (see below) is potential inside the substrate, and $V(r) = 0$ inside the quantum dot. The electron effective mass $m^* = m^*(x, y, z, E)$ is linearly dependent on energy for $0 < E < V_c$ and varies within the limits of the QD/substrate bulk effective mass values [3]. $V_c$ is defined as $V_c = \kappa \left( E_{g,2} - E_{g,1} \right)$, where $E_g$ is the band gap and the coefficient $\kappa < 1$ is different for conduction and valence bands. We use $\kappa^{CB} = 0.42$ and $\kappa^{VB} = 0.58$, Ref. [2]. For experimental values $E_{g,1} = 1.1$ eV and $E_{g,2} = 8.9$ eV, the band gap potential for the conduction band (valence band) is $V_c = 3.276$ eV ($V_c = 4.524$ eV). Bulk effective masses of Si and SiO$_2$ are $m^*_{0,1} = (m_{0,1\perp} = 0.19\, m_0$, $m_{0,1\parallel} = 0.91\, m_0)$ and $m^*_{0,2} = 1.0\, m_0$ respectively, where $m_0$ is the free electron mass. The value of $m^*_{0,1} = 0.34\, m_0$ is used for the effective mass of the heavy hole for Si QD and $m^*_{0,2} = 5.0\, m_0$ - for substrate. The presented values applied as initial parameters for the Si/SiO$_2$ QD may be changed due to the effective mass dependence [3]. The band structure of the model is shown in Fig. 1a). Energy dependence of electron effective mass $m_{0,1\perp}$ is shown in Fig. 1b).

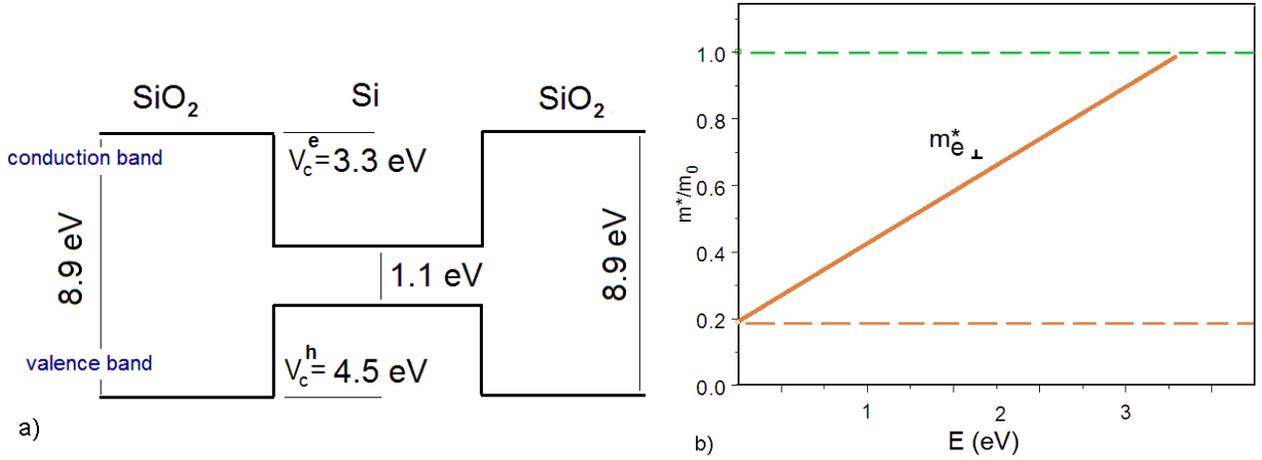

Fig. 1 a) Band structure for Si/SiO$_2$ QDs. b) Energy dependence of electron effective mass. Bulk values of effective mass in Si and SiO$_2$ are given by the dashed lines.

Note that the charge image potential, strain-induced potential, and inter-band interaction are not taken into account in this model. We applied the finite element method to find numerical solution of Eq. (1).

Since the problem may be considered as one having rotational symmetry cylindrical coordinates are used to formulate mathematical problem for the numerical solution.

## 3. Small size QD

Calculations of ground state electron/hole energy levels are performed for the case of small size (diameter $D \leq 6\,\mathrm{nm}$) QDs. For the Si/SiO$_2$ heterostructure the PL experimental data [4-7] are available for comparison with our theoretical considerations. We used first order perturbation theory to calculate neutral exciton recombination energy taking into account the Coulomb force between electron and heavy hole. The results of our calculation are shown along with the experimental data [4-7] in Fig. 2.

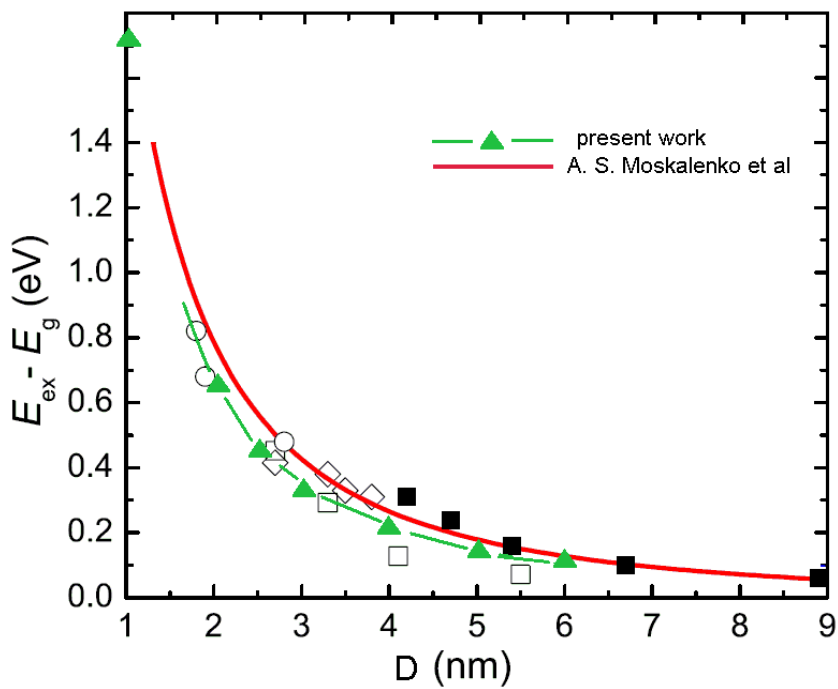

Fig. 2 Neutral exciton recombination energy $E_{ex}$ of Si/SiO$_2$ spherical QDs. Calculated values are shown by triangles. Solid curve is result of [2]. Experimental data are taken from [1, 4-7] and are shown by various symbols.

The PL exciton data are reproduced well by our model. We also compare the results with those obtained within the model given in [2]. The comparison shows that inter-band interaction for heavy and light holes taken into account in [2] has character of second order term and cannot explain the difference with our results. We found that the Coulomb shift of exciton energy is slightly larger than one evaluated in [2]. In our calculations the energy dependence of the electron/hole effective mass [3] is used to take into account the non-parabolicity of conduction/valence band. This assumption affects the essential change of the effective mass for the carriers from the initial values chosen to be equal to the bulk values for the materials of QD and substrate. The results are shown in Fig. 2. The energy levels of confined electron/hole are slightly shifted to the bottom of the conduction/valence band from the values calculated under the parabolic approximation. It causes a shift of our results to smaller values with respect to the results of the parabolic calculation [2] in Fig. 3. We have found that the Coulomb shift of the exciton energy is slightly larger than one evaluated by the formula $U_C = -0.3689\, D^{-1}$ in [2]. The result obtained in present work can be fitted by $U_C = -0.3244\ D^{-\alpha}$, $\alpha = 0.886$.

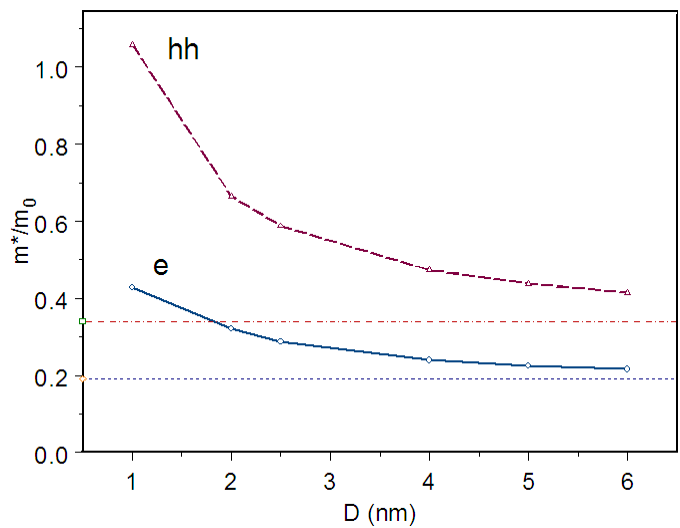

Fig. 3 Change of effective mass of the carriers is shown along diameter of QDs.

## 4. Weak confinement regime

For weak confinement regime (Si/SiO$_2$ QD diameter $D \geq 10$ nm), when the number of confinement levels is of the order of several hundred, we studied neighboring level statistics of the electron/hole spectrum. Generally the effects of mass anisotropy and shape of quantum dots are assumed as possible sources of chaos in electron spectrum of QD. It is well known that the anisotropic Kepler problem shows the strong chaotic properties [8] which, in particular, lead to the Wigner-Dyson-like or GOE-like Random Matrix Theory (RMT) behavior [9]. The Si quantum dots having strong difference of electron effective mass in two directions is considered appropriate example for the study of role of the effective mass asymmetry. In this study we do not include the Coulomb potential between electrons and holes. The shape geometry role is studied for two and three dimensions.

The low-lying single electron/hole energy levels are notated by $E_i$, $i = 0,1,...N$. One can obtain the set $\Delta E_i = E_i - E_{i-1}$, $i = 1,...N$ of energy differences between neighboring levels. Our goal is an evaluation of the distribution function $R(\Delta E)$ of differences of the neighboring levels. The function is normalized by $\int R(\Delta E) d\Delta E = 1$. For numerical calculation we define a finite-difference analog of the distribution function:

$$R_j = N_j / H_{\Delta E} / N, \; j = 1,...M ,  \qquad (2)$$

where $\sum N_j = N$ represents total number of considered levels, $H_{\Delta E} = ((\Delta E)_1 - (\Delta E)_N)/M$ is the energy interval which we obtained by dividing the total region of energy differences by M bins. $N_j$ ( $j = 1,...M$ ) is the number of energy differences which are located in the $j$ -th bin.

At the beginning we considered quantum well (QW) with infinite walls in two dimensions (2D): $\left( -(\nabla, \frac{\hbar^2}{2m^*} \nabla) + V(r) - E \right) \psi = 0$, where $V(r) = 0$ in QW and $V(r) = \infty$ everywhere outside of QW. When the boundaries of the QW are rectangular the energy levels are expressed as $E_{n,m} = a(n^2 + m^2 / v), n = 1,2,...,m = 1,2..., v = m_\parallel^* / m_\perp^*, a = const$. For this case distribution of the neighboring levels differences is of course Poisson-like as we found. For the case of nonrectangular shape of QW one can obtain a non Poisson distribution for $\Delta E$ by choosing the shape with hyperbolic boundaries. This result is shown in Fig 4a. In Fig. 4b we show the shape of this QW. The curve of Fig. 4a (and all figures below) is obtained by the spline smoothing method for calculated values $R_j$ (see Eq. (2)). In Fig. 4a the calculated values $R_j$ are shown by histogram[*].

Role of QD geometry with different shapes of boundaries is studied in details previously (see e.g. [11]). Our results for the distribution function of the ellipsoidal shaped Si/SiO₂ QD are presented in Fig. 5a. In the inset figure we show the cross section of the QD. The fitting of the calculated points for $R(\Delta E)$ gives the Poisson-like distribution. For the case of QD shape with the break of the ellipsoidal symmetry (Fig. 5b) by the cut below the major axis we obtained a non-Poisson distribution for $\Delta E$ like for 2D QW shown above (Fig. 4a). For the ellipsoid with the cut (Fig. 5b) one can obtain fit for $R(\Delta E)$ using the Brody distribution [10]:

$$R(s) = (1 + \beta) b s^{\beta} \exp(-b s^{1+\beta}), \tag{3}$$

with the parameter $\beta = 1.0$ and $b = (\Gamma[(2 + \beta)/(1 + \beta)]/H)^{1+\beta}$, $H$ is the average level spacing. Note that for the Poisson distribution the Brody parameter is equal zero.

---

[*] In the following Figures 5-8, we do not show histograms, they are replaced by smooth curves.

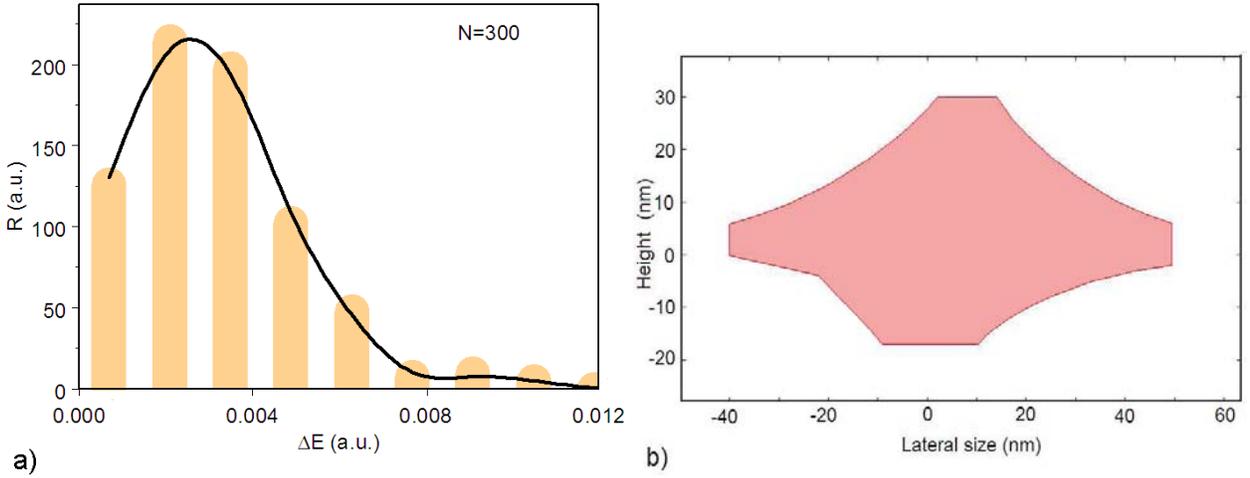

a) b)

Fig. 4 a) Distribution functions for electron energy levels $R(\Delta E)$ of 2D QW ($\frac{\hbar^2}{2m^*} = 1/2$). Number of electron levels is 300. The Brody parameter for the curve which fits the distribution [10] are $\beta = 0.82$ in Eq. (3). b) The QW shape area.

It is worthy to remark that, in principle, the quantum chaos with its level repulsion and RMT are not in one-to-one correspondence. It was known long ago that the classically chaotic system with geometry of negative curvature can have spectral properties inconsistent with RMT [12].

The Poisson and RMT statistics do not exhaust all possible behaviors, and there are spectral statistics with the properties intermediate between their regimes [13-14] (see also review [15]). There exist so called hybrid statistics for some dynamical systems which are neither integrable nor chaotic and belong to the class of pseudo-integrable systems [16] (see Fig. 6 for polygon-like shape QD). Their chaotic character (expressed, e.g., by the level repulsion) results not from the divergence of the close trajectories but from the bifurcation of beams of trajectories hitting the polygon vertices [16]. These hybrid statistics reveal the level repulsion at small spacings as in RMT and have Poisson-like increase at large spacings [17].

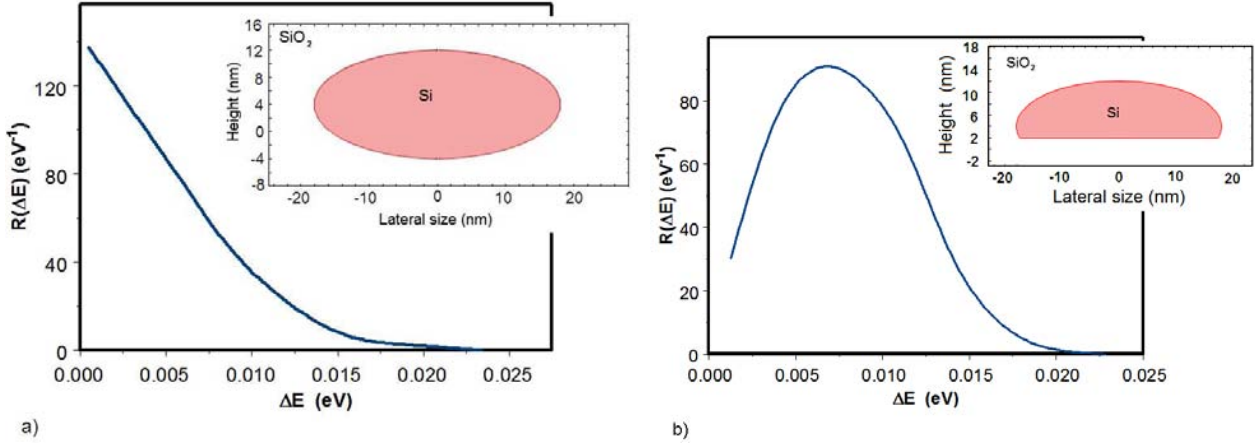

Fig. 5 Distribution functions for electron neighboring levels in Si/SiO$_2$ QD for different shapes: a) ellipsoidal shape, b) ellipsoidal like shape with cut. The Brody parameter $\beta$ is equal to 1 in Eq. (3). Statistics for $s$-shell levels are considered. In the inset the QD cross section is shown.

Figures 7 and 8 give examples of non-Poisson distribution for the QD with shapes having the breaking of discrete symmetry form. In Fig. 7 the spherical symmetry (giving the Poisson statistics) is changed by semi-spherical cavity. Change in statistics is indicated by the distribution function of neighboring levels shown here. In Fig. 8 we present another variant when the Poisson statistics is restored by restoring the shape reflection symmetry. In this connection we would like to mention recent work [18] (see also [19]) where the reflection symmetry arguments were used to predict an enhanced constructive interference effect in the conductance of the double-dot structures.

Phenomena observed in the present paper looks like the expression of the above described complicate and sometime non- universal situation [17, 20]. Particularly, in [20] it is suggested that there exist quantum systems with no discrete symmetry and with well - defined classical limit with chaotic dynamics, which, nonetheless, does not have RMT level statistics.

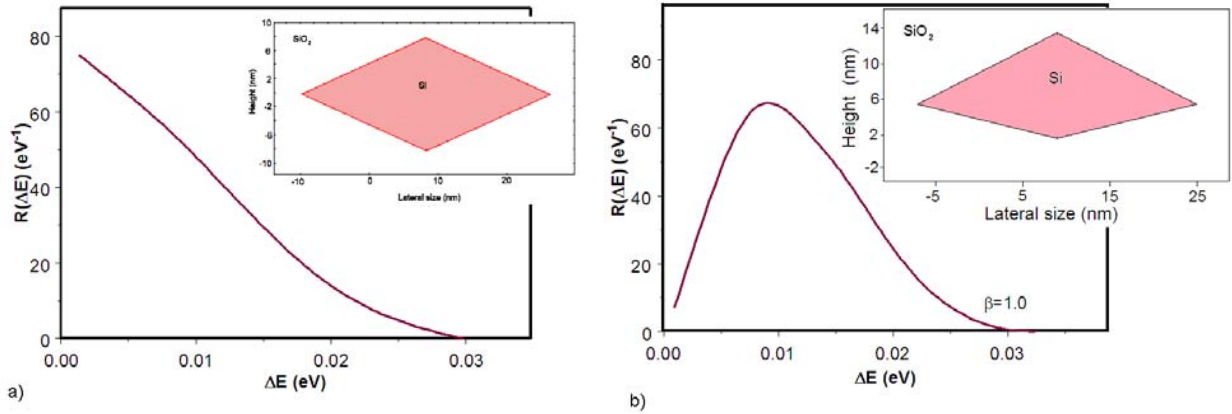

Fig. 6 Distribution functions for electron neighboring levels in Si/SiO₂ QD for polygon-like shapes: a) symmetric rhomb and b) slightly deformed rhomb (without discrete symmetry) as cross section of the QD shape. In insets the QD cross sections are shown.

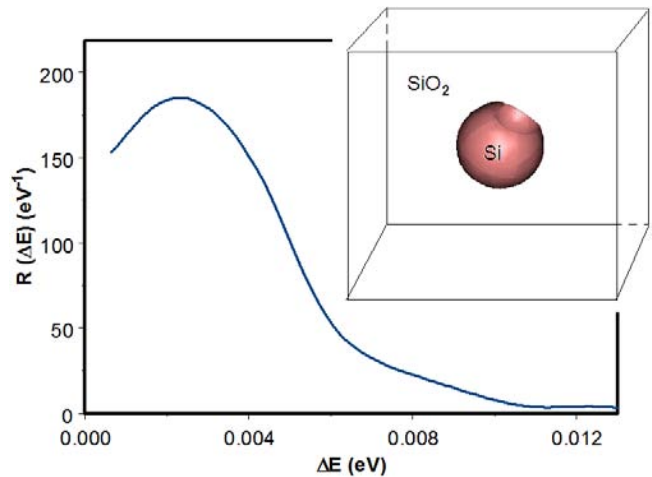

Fig. 7 Distribution functions for electron neighboring levels in Si/SiO₂ QD for spherical-like shape with cut. In inset the geometry of this QD are shown in 3D, the QD diameter is 17 nm. The Brody parameter $\beta$ is equal to 1.0 in Eq. (3).

## 5. Conclusion

We found that the PL spectra of neutral exciton recombination energy of Si/SiO$_2$ spherical QDs can be well described by our model. The effect of non-parabolicity is important for the small size QDs.

We found that the effective mass anisotropy in Si/SiO$_2$ QDs does not lead to the essential influence on the neighbor levels distribution of single electron/hole. In the same time the deformations of shape of QD strongly affect the statistical properties of QD energy levels, and it may has technological implications. In particular, the deviation of the shapes from the symmetric ones leads to the non-Poisson statistics, which may be a sign of the quantum chaos. This, particularly, will influence the conductance and other transport properties of the QD.

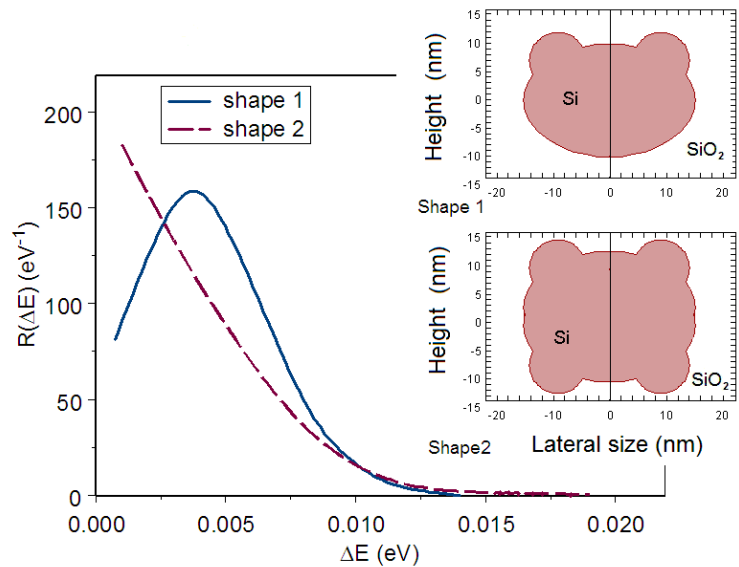

Fig. 8 Distribution functions for electron neighboring levels in Si/SiO$_2$ QD for different elliptical-like shapes. In insets the QD cross sections are shown. The Brody parameter $\beta$ for distribution of "shape 1" is equal to 0.95 in Eq. (3).

## Acknowledgments

This work is supported by NSF CREST award HRD-0833184 and NASA award NNX09AV07A.